\documentclass[twocolumn,showpacs,preprintnumbers,amsmath,amssymb]{revtex4}

\usepackage{graphicx}
\usepackage{dcolumn}
\usepackage{bm}

\begin{document}

\preprint{APS/123-QED}

\title{Surface-enhanced optical third-harmonic generation in Ag island films}

\author{ E.M. Kim, S.S. Elovikov, T.V. Murzina, A.A. Nikulin}
\author{O.A. Aktsipetrov}%
\email{aktsip@shg.ru}

\affiliation{%
Department of Physics, Moscow State University, 119992 Moscow,
Russia}%
\author{M.A. Bader and G. Marowsky}
\affiliation{%
Laser-Laboratorium G\"ottingen, D-37077 G\"ottingen, Germany}%
\date{\today}

\begin{abstract}
{Surface-enhanced optical third-harmonic generation (THG) is
observed in silver island films. The THG intensity from Ag
nanoparticles is enhanced by more than two orders of magnitude
with respect to the THG intensity from a smooth and homogeneous
silver surface. This enhancement is attributed to local
plasmon excitation and resonance of the local field at the third-harmonic
wavelength. The diffuse and depolarized component of the enhanced THG
is associated with the third-order hyper-Rayleigh scattering in a
2-D random array of silver nanoparticles.}
\end{abstract}

\maketitle Observation of surface-enhanced nonlinear optical
effects in silver island  films dates back to two 1981 papers by
Wokaun, et al., [1,2], where surface-enhanced optical second-harmonic
generation (SHG) and  surface-enhanced Raman scattering (SERS)
were observed in silver island films. The enhancement of the SHG
intensity by up to three orders of magnitude  was attributed in
[1] to the resonant enhancement of the local field  at the
second-harmonic (SH) wavelength, mediated by the excitation of the
local surface plasmons in silver nanoparticles. This plasmon
mechanism of the local field enhancement, introduced by Berreman
[3] and Moskovits [4], was intensively discussed in the context of SERS active
structures and surface-enhanced SHG from electrochemically
roughened silver surfaces [5] and rough surfaces of
other metals [6]. According to this approach, nonlinear
polarization of a rough metal surface or an array of small metal
particles  is given by:
$P_{2\omega}=L_{2\omega}\chi^{(2)}(2\omega)L_{\omega}^{2}
E_{\omega}^{2}$ at the SH wavelength and
$P_{3\omega}=L_{3\omega}\chi^{(3)}(3\omega)L_{\omega}^{3}E_{\omega}^{3}$
at the third-harmonic (TH) wavelength, where $\chi^{(2)}(2\omega)$
and $\chi^{(3)}(3\omega)$ are the second- and third-order
susceptibilities of metal, respectively; $E_{\omega}$ is the
optical field at fundamental wavelength; and $L_{\omega}$,
$L_{2\omega}$ and $L_{3\omega}$ are the local field factors at the
corresponding wavelengths.

The spectral dependence of the local field factor of an array of
small metal spheroids embedded in a dielectric matrix within the
simple approach in dipole and effective media approximations, is
given by [7]:
\begin{equation}
\displaystyle
L(\lambda)=\frac{\varepsilon_{d}(\lambda)}{\varepsilon_{d}(\lambda)
+[\varepsilon_{m}(\lambda)-\varepsilon_{d}(\lambda)](N-q/3)},
\label{1}
\end{equation}
where $\varepsilon_{d}(\lambda)$ and $\varepsilon_{m}(\lambda)$
are the dielectric constants of the dielectric matrix  and of the
metal, respectively; $N$ is the shape-dependent depolarization
factor of the spheroids; and $q$ is the filling factor, i.e., the
relative fraction of the metal in a composite material. The
resonant wavelength of the local field factor, $\lambda_{res}$,
corresponds to setting the real part of the denominator in Eq. 1 to zero:
$Re[{\varepsilon_{d}(\lambda_{res})
+[\varepsilon_{m}(\lambda_{res})-\varepsilon_{d}(\lambda_{res})](N-q/3)]=0}$.
For an isolated small Ag sphere in vacuum $\lambda_{res}\approx
200 nm$. Three factors result in the red-shift of $\lambda_{res}$
up to the visible range for an array of particles: (1) the distortion of
the particle shape, (2) the dipole-dipole interaction between
particles and, (3) an increase in the dielectric constant of the matrix
material. The resonant increase of the local field factor at
$\lambda_{res}$ results in a many-fold increase of the
nonlinear-optical response from a nanoparticle array.

Up to the present time, the experimental studies of local plasmon
enhancement in island films were restricted to SERS and SHG. Fig. 1d,
taken from [1], shows the dependencies of the
local field factor in Ag island films on mass thickness,
$d_{m}=m/\rho$, where $m$ is the mass of metal deposited per unit
area and $\rho$ is the bulk density of Ag. These dependencies
show the maximum at $d_{m}$=2.0 nm and $d_{m}$=6.0 nm for
wavelengths of 532 nm and 1064 nm, respectively. The decrease of
$d_{m}$ results in the blue-shift of $\lambda_{res}$ due to a
decrease of interparticle interaction. One might anticipate that the
resonance at THG wavelength $\lambda_{res}$ = 355 nm for the 1064 nm
fundamental wavelength can be attained for $d_{m}\sim 1 nm$.

Another peculiarity of the surface-enhanced SHG from metal island
films is the diffuseness of the SH radiation [5,6]. This is a
manifestation of the second-order hyper-Rayleigh scattering.
Silver island films are random arrays of nanoparticles which
possess random spatial inhomogeneity of nonlinear susceptibilities
[8] and local field factors [9]. This inhomogeneity is the source
of the diffuse SHG radiation in hyper-Rayleigh scattering.

The enhancement and diffuseness observed, up to now, for SHG
are general features of the nonlinear optical effects in island
films and are supposed to be observed in THG. Meanwhile, in spite
of this analogy, there is a principle difference between SHG and THG
in metal nanoparticles: $\chi^{(2)}$ is localized at the surface
of nanoparticles and vanishes in the bulk of a centrosymmetric
metal, whereas $\chi^{(3)}$ is a bulk localized nonlinearity [10],
as shown schematically in the inset in Fig. 1a.

In this paper, surface-enhanced THG and third-order hyper-Rayleigh
scattering is observed in Ag island films. The resonant plasmon
mechanism of the THG enhancement is proved.

\begin{figure}[!h]
\vspace{1 cm}
\begin{centering}
\includegraphics[width=8cm]{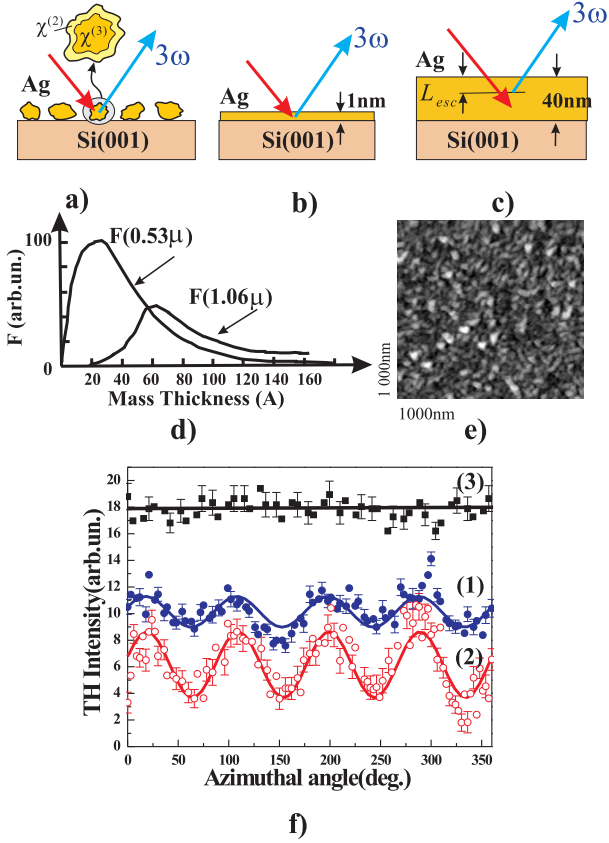}
\caption{Schematics of a) silver island film (inset in (a) shows
the localization of $\chi^{(2)}$ and $\chi^{(3)}$ in
nanoparticles); b) model of homogeneous film of the equivalent
thickness of 1 nm; c) homogeneous reference Ag film with a
thickness of 40 nm and an escape depth of $L_{esc}=7 nm$ for the
TH wave at the fundamental wavelength $\lambda_{\omega}= 1064 nm$;
d) dependencies of the local field factors at wavelengths of 1.064
$\mu$m and 0.532 $\mu$m on the mass thickness (from [1]); e) the
AFM image of silver island film; f) the THG intensity as a
function of the azimuthal angle from: (1) silver island film on
Si(001) substrate, (2) silver-free Si(001) substrate, (3)
homogeneous reference Ag film with a thickness of 40 nm; the solid
lines are results of approximations.} \label{1}
\end{centering}
\vspace{1 cm}
\end{figure}

The films were prepared by thermal evaporation of silver onto the
substrates of silicon Si(001) wafers at a rate of 3-4 $\AA/s$ and
residual pressure of $10^{-5}$ Torr. The silicon wafers were
chosen as substrates because of, first, the flatness and
homogeneity of the surface, and, second, the simplicity of the
chemical etching procedure that is used for the preparation of a
step-like SiO$_2$ wedge (see the scheme in Fig. 2b). Three types
of Ag films are studied: Ag island film with a mass thickness of
$d_{m}\approx$ 1 nm and expected plasmon resonance at
$\lambda_{res}\sim$ 355 nm, Ag island film with plasmon resonance
in the vicinity of $\lambda_{res}\sim$ 270 nm on the silicon oxide
step-like wedge, and a thick homogeneous Ag reference film with a
thickness of 40 nm. The thick homogeneous Ag film is used as a
reference source of non-enhanced bulk THG for the measurement of
the THG enhancement from island films. An atomic force microscopy
(AFM), in the constant force mode, and with a height resolution of
1 nm and lateral resolution of approximately 10 nm, is used to
characterize the morphology of the samples. Fig. 1e shows the AFM
image of Ag island film. A cross-section of the profile shows that
the average lateral  size and height of silver nanoparticles is
about 40 nm and 3 nm, respectively.

The outputs of two laser systems are used as the fundamental
radiation in the THG and SHG experiments: (1) an OPO laser system,
"Spectra-Physics ÌÎÐÎ 710," with a wavelength which is tunable in the
spectral range from 490 nm to 680 nm, a pulse duration of 4 ns, and a pulse
intensity of 2 MW/cm$^2$; and (2) a Q-switched YAG:Nd$^{3+}$ laser tuned to a
1064 nm wavelength, a pulse duration of 15 ns, and a pulse
intensity of about 1 MW/cm$^2$. The TH(SH) radiation is filtered out
by appropriate UV and BG color and bandpass filters and detected by
a PMT and gated electronics. To normalize the THG(SHG) intensity
over the OPO and YAG:Nd$^{3+}$ laser fluency, and the spectral
sensitivity of the optical detection system, a reference channel is
used with a Z-cut quartz plate as a nonlinear optical reference and with a
detection system identical to that of the "sample" channel. Polar
rotation of the detector system enables us to measure the linear
Rayleigh scattering pattern and the THG and SHG scattering patterns (see
Fig. 2a).

\begin{figure}[!h]
\vspace{1 cm}
\begin{centering}
\includegraphics[width=7.8cm]{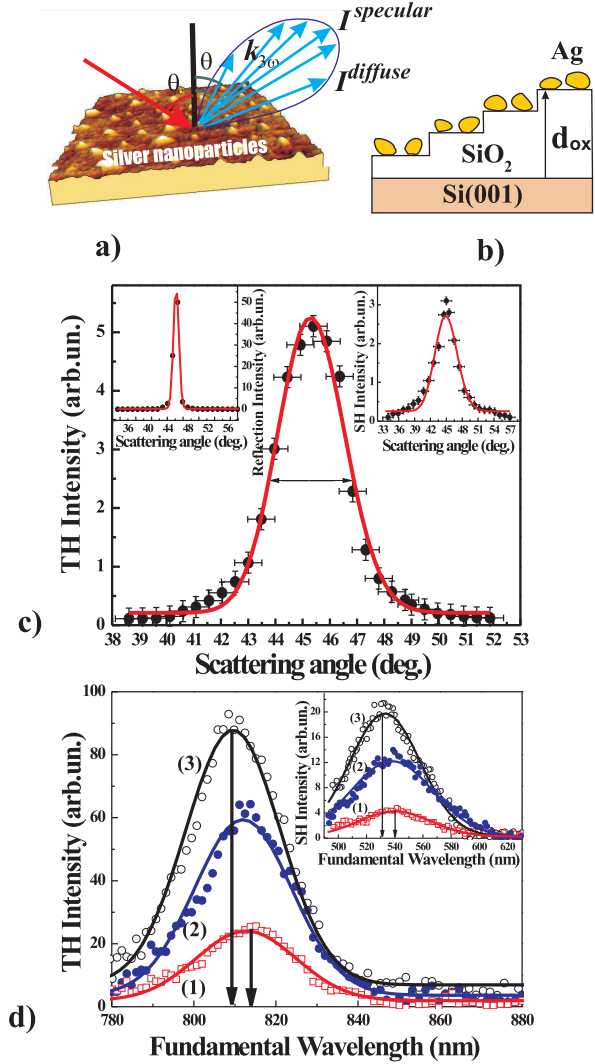}
\caption {a) The scheme of the hyper-Rayleigh scattering
experiment; b) scheme of silicon oxide step-like wedge as a
variable spacer between the island film and the high-dielectric
constant material (silicon); $d_{ox}$ is the thickness of the
silicon oxide steps; c) THG scattering pattern: the THG intensity
from the island film as a function of polar scattering angle
(angular width is  $3^{\circ}\pm 0.5^{\circ}$); left inset is the
linear Rayleigh scattering pattern, with angular width of
$1^{\circ}\pm 0.5^{\circ}$; right inset is the SHG scattering
pattern, with angular width of $5^{\circ}\pm 0.5^{\circ}$;  solid
lines are approximations with Eq. 3; d) main panel: dependencies
of the THG intensity on the fundamental wavelength for Ag island
films deposited on the SiO$_2$ step-like wedge for $d_{ox}$= 2 nm
(1), 70 nm (2) and 100 nm (3); inset: the same for the SHG
intensity spectra; solid lines are approximations by Gaussian
function.} \label{2}
\end{centering}
\vspace{1 cm}
\end{figure}

In order to observe and measure an enhancement in  THG, two
experimental points have to be taken into account: (1) the THG
signal from Ag island film should be distinguished from the
Si(001) substrate contribution and (2) the THG intensity should be
integrated over the diffuse THG scattering pattern. The following
paragraphs focus on these points.

The curve (1) in Fig. 1f shows the azimuthal dependence of the THG
intensity from the sample of Ag island film in the specular
direction for the ${\it s}$-in, ${\it s}$-out combination of
polarizations of the fundamental and TH waves. The anisotropic
component of the THG signal is related to the nonlinear response
of Si(001) substrate, whereas the isotropic THG is attributed to
both Ag nanoparticles and Si substrate. To distinguish the THG
contribution of Ag island film from that of the substrate, the
azimuthal dependence of the THG intensity from Si(001) is measured
in the same ${\it s}$-in, ${\it s}$-out geometry  (curve (2) in
Fig. 1f). The ratio of the anisotropic components of THG from
silver island film on Si(001), $I^{anis}_{IF+Si}(3\omega)$, and
silver-free Si(001) substrate, $I^{anis}_{Si}(3\omega)$, gives an
attenuation coefficient of the THG response from substrate due to
the absorption and scattering in the silver coverage:
$\displaystyle
\alpha_{3\omega}=I^{anis}_{IF+Si}(3\omega)/I^{anis}_{Si}(3\omega)=0.46$.
The estimation of the $\displaystyle \alpha_{3\omega}$ coefficient
is the first step in the determination of the THG enhancement for
Ag island films. The relative value of the THG intensity from Ag
nanoparticles in the specular direction is
$I_{IF}^{spec}(3\omega)=I^{is}_{IF+Si}(3\omega)-\alpha_{3\omega}I^{is}_{Si}(3\omega)$,
where $I^{is}_{IF+Si}(3\omega)$ and $I^{is}_{Si}(3\omega)$ are the
isotropic components of THG from the island film on Si(001) and
from silver-free substrate, respectively. Obtaining the total
intensity of the diffuse THG from Ag nanoparticles is the second
step in the determination of the THG enhancement. This demands the
measurement of the THG scattering pattern which is dependent on
the diffuse THG intensity, which in turn depends on the polar
scattering angle:
$I_{IF}(3\omega,\theta)=I_{IF}^{spec}(3\omega)S_{3\omega}(\theta)$,
where $\theta$ and $S_{3\omega}(\theta)$ are the polar scattering
angle and the normalized form-factor of the third-order
hyper-Rayleigh scattering, respectively. The main panel in Fig. 2c
shows the experimental scattering pattern of diffuse THG from
silver island films where the angular width of the normalized
form-factor is approximately $3^{\circ}\pm 0.5^{\circ}$. This
sufficiently exceeds the angular width of $1^{\circ}\pm
0.5^{\circ}$ of the scattering pattern of linear Rayleigh
scattering from the same silver island film presented in the left
inset in Fig. 2c, whereas the angular width of the SHG scattering
pattern is approximately $5^{\circ}\pm 0.5^{\circ}$. The total
intensity of the diffuse THG can be obtained by angular
integration of the THG scattering pattern and in the case of the
small angular width of $S_{3\omega}(\theta)$, is given by:
$I_{IF}(3\omega)=\displaystyle I_{IF}^{spec}(3\omega)
[\frac{1}{\Omega}\int_{\Delta \theta}
S_{3\omega}(\theta)d\theta]^{2}\approx 0.6 \cdot 10^{2}
I_{IF}^{spec}(3\omega)$, where ${\Omega}$ is an angular aperture
of the THG detection system and $\Delta \theta$ is the angular
interval of integration. To estimate quantitatively the THG
enhancement, we consider the total THG intensity of the specular
THG from a model homogeneous film with an equivalent thickness of
$d_{m}$=1 nm and compare this with the THG intensity from a
reference film with a thickness of 40 nm (see Fig. 1b and c). The
THG intensity detected from  the reference film comes from the Ag
layer corresponding to the escape depth $L_{esc}$ of the TH wave
(curve (3) in Fig. 1f). In our experimental conditions,
$L_{esc}\sim$ 7 nm for $\lambda_{3\omega}$= 355 nm. Thus, the
enhancement of the THG intensity from Ag island film with respect
to a thick homogeneous Ag film is given by:
\begin{equation}\displaystyle G=\frac{I_{IF}(3\omega)}{I_{ref}(3\omega)}\left [\frac {\int_0^{40}
[\exp-(\alpha_{3\omega}+3\alpha_{\omega})r]dr}{\int_0^{1}
[\exp-(\alpha_{3\omega}+3\alpha_{\omega})r]dr}\right]^{2} \approx
1.2\cdot 10^2, \label{3}
\end{equation}
where $I_{ref}(3\omega)$ is the THG intensity from reference Ag
film, $\alpha_{\omega}$ and $\alpha_{3\omega}$ are adsorption
coefficients at the fundamental and TH wavelengths, respectively,
and $r$ is coordinate normal to the film surface.

To prove the plasmon assistance in the THG enhancement, the
dependencies of the THG intensity on the fundamental wavelength are
studied for Ag island films deposited onto a step-like SiO$_2$
wedge on silicon wafer. Silicon oxide steps serve as variable
spacers between the silver nanoparticles and silicon substrate,
which is a high-dielectric constant material. Variations of
SiO$_{2}$ thickness, $d_{ox}$, result in the variations of the
effective dielectric constant
$\varepsilon_{dl}(d_{ox})=\varepsilon_{dl}^{\prime}(d_{ox})+ i
\varepsilon_{dl}^{\prime \prime}(d_{ox})$ in Eq. 1: the increase of
$d_{ox}$ corresponds to the decrease of the effective
$\varepsilon_{dl}(d_{ox})$. Theoretical modeling [11] shows that
the decrease of the real part,
$\varepsilon^{\prime}_{dl}(d_{ox})$, results in the blue-shift of
the resonant plasmon wavelength, whereas the decrease of the
imaginary part, $\varepsilon^{\prime \prime}_{dl}(d_{ox})$, leads
to the enhancement of the local field amplitude.

The main panel in Fig. 2d shows a set of THG spectra for $d_{ox}$
increasing in the range from 2 nm to 100 nm. The observed effects of $d_{ox}$
on the THG spectra as SiO$_2$ thickness increases from 2 nm to 100 nm
is two-fold: (1) an apparent blue-shift of approximately 6 nm of the
THG resonance and, (2) a more than four-fold increase of the THG
resonant intensity. These changes correspond to the decrease of the
effective dielectric constant of Ag islands situated at different
steps of the SiO$_{2}$ wedge. An analogous blue-shift of about 10
nm and manyfold increase of the resonant SHG intensity are observed
in the same conditions (refer to the inset in Fig. 2d). The impact
of the dielectric constant of the substrate on the resonant properties
of surface-enhanced THG and SHG proves the plasmon-assisted mechanism
of the enhancement. A slight difference in the spectral shift for THG
and SHG can be associated with the different localizations of
$\chi^{(3)}$ and $\chi^{(2)}$ in metal nanoparticles. Moreover,
different spatial localizations of nonlinear susceptibilities in
metal particles can result in different parameters of scattering
patterns at the TH and SH wavelengths from a random array of Ag
nanoparticles. Normalized form-factors at the TH and SH wavelengths
are given by:
\begin{equation}
\displaystyle S_{3\omega, 2\omega}(\theta)\sim \exp[-M_{3\omega,
2\omega}k_{3\omega, 2\omega}^{2}l^{2}_{cor}(3\omega, 2\omega)],
\label{3}
\end{equation}
where $l_{cor}(3\omega, 2\omega)$ is the correlation length at the
TH and SH wavelengths, respectively, $\displaystyle k_{3\omega,
2\omega}=2\pi(sin\theta-sin\theta_{0})/\lambda_{3\omega, 2\omega}$
and $M_{3\omega, 2\omega}$ is an adjustable parameter at the TH
and SH wavelengths, respectively; and $\theta_{0}$ is the angle of
incidence. The approximation of the diffuse THG and SHG scattering
patterns (Fig. 2c, solid lines) by Eq. 3 corresponds to the
correlation lengths of $l_{cor}(3\omega)\approx$ 42 nm and
$l_{cor}(2\omega)\approx$ 20 nm. The former probably corresponds
to the average Ag particle size of 40 nm obtained from the AFM
measurements because of the bulk localization of $\chi^{(3)}$.
Meanwhile, the latter being twice as small as $l_{cor}(3\omega)$,
corresponds to the smaller scale of the $\chi^{(2)}$ inhomogeneity
due to its surface localization within the individual particles.

In conclusion, surface-enhanced THG is observed in Ag island films
with an enhancement of $1.2\cdot10^{2}$, which is attributed to
the local surface plasmon excitation in Ag nanoparticles at the TH
wavelength. Diffuseness of surface-enhanced THG allows us to
associate this effect with the third-order hyper-Rayleigh
scattering in a random array of nanoparticles. The difference in
scattering patterns and spectroscopic resonances between
surface-enhanced THG and SHG can be attributed to the different
localizations of $\chi^{(3)}$ and $\chi^{(2)}$ in nanoparticles.

\begin{acknowledgments}
This work is supported in part by the Russian Foundation for Basic
Research (Grants No. 04-02-16847, 04-02-17059, 03-02-39010), the
Presidential Grants for Leading Russian Science Schools (No.
1604.2003.2 and 1909.2003.2), the DFG Grant No. 436 RUS
113/640/0-1, the NATO Grant PST.CLG.979406 and the INTAS Grant No.
03-51-3784.
\end{acknowledgments}


\begin{thebibliography}{99}

\bibitem {c1} A. Wokaun, J.G. Bergman, J.P. Heritage, A.M. Glass, P.F.
Liao and D.H. Olson, Phys. Rev. B. {\bf 24}, 849 (1981).

\bibitem {c2} A. Wokaun, J.P. Gordon and P.F.
Liao, Chem. Phys. Lett. {\bf 48}, 957 (1981).

\bibitem {c3} D.W. Berreman, Phys. Rev. {\bf 163}, 855 (1967).

\bibitem {c4} M. Moskovits, J.Chem. Phys. {\bf 69}, 4159 (1978).

\bibitem {c5} C.K. Chen, A.R.B. de Castro and Y.R. Shen,
Phys. Rev. Lett. {\bf 46}, 145 (1981).

\bibitem {c6} G.T. Boyd, Th. Rasing, J.R.R. Leite and Y.R. Shen, Phys. Rev. B. {\bf 30}, 519 (1984).

\bibitem {ñ7} V.I. Emelyanov, N.I. Koroteev, Soviet Fisiks-Uspekhi, {\bf 135}, 345 (1981).

\bibitem {c8}O.A. Aktsipetrov, P.V. Elyutin, A.A. Nikulin and E.A. Ostrovskaya, Phys. Rev. B.{\bf 51}, 17 591-599, (1995).

\bibitem {c9}  M. Breit, V.A. Podolsky, S. Gresillon, G. von Plessen, J. Feldman, J.C. Rivoal,
 P. Gadenne, V.M. Shalaev, A.K. Sarychev, Phys. Rev. B.{\bf 64}, 125106-1, (2001).
\bibitem {c10}  N. Bloembergen, R.K. Chang,  S.S. Jha, C.H. Lee, Phys. Rev.{\bf 147}, 813-822, (1968).

\bibitem {c11} O.A. Aktsipetrov, E.M. Dubinina, S.S. Elovikov, E.D. Mishina, A.A. Nikulin, N.N. Novikova, M.S.
Strebkov, Sol. Stat. Comm.{\bf 70}, 1021-1024, (1989).




\end{thebibliography}
\end{document}